\documentstyle[multicol,aps,psfig]{revtex}
\renewcommand{\narrowtext}{\begin{multicols}{2}
\global\columnwidth20.5pc\noindent}
\renewcommand{\widetext}{\end{multicols}
\global\columnwidth42.5pc}
\begin{document}
\draft
\preprint{October 27, 2000}
\title{
Coexistent quantum and classical aspects of magnetization plateaux
in alternating-spin chains
}
\author{T$\hat{\mbox o}$ru Sakai}
\address
{Faculty of Science, Himeji Institute of Technology,
 Ako, Hyogo 678-1297, Japan}
\author{Shoji Yamamoto}
\address
{Department of Physics, Okayama University,
 Tsushima, Okayama 700-8530, Japan}
\date{\today}
\maketitle
\begin{abstract}
Magnetization process of ferrimagnetic Heisenberg chains of alternating
spins are theoretically studied.
The size scaling analysis with the exact diagonalization of finite
systems for ($S$,$s$)=(3/2,1) and (2,1) indicates
a multi-plateau structure in the ground-state magnetization
curve for $S$ and $s$ $>1/2$.
The first plateau at the spontaneous magnetization can be explained by a
classical origin, that is the Ising gap.
In contrast, the second or higher one must be originated to the
quantization
of the magnetization.
It is also found that all the $2s$ plateaux, including the classical and
quantum ones, appear even in the isotropic case with no bond
alternation.
\end{abstract}
\pacs{PACS numbers: 75.10.Jm, 75.40Mg, 75.30.Kz}
\narrowtext

\section{Introduction}

Alternating spin chains with antiferromagnetic interactions have 
been attracting a lot of current interest. 
They behave like a gapped antiferromagnet based on the optical 
mode of the low-lying excitation, 
while they also exhibit a ferromagnetic aspect 
characterized by a spontaneous magnetization. 
The coexistence of the two aspects gives rise to various interesting crossover 
phenomena at low temperatures.(Yamamoto 2000) 
However, few {\it quantum} aspects have ever been reported for the systems. 
Then we consider an interesting phenomenon caused 
by a quantum mechanism, called {\it 
quantization of magnetization}, 
on the stage of the quantum ferrimagnetic chains. 
It would be observed as a plateau in the ground-state magnetization curve. 
Recently many theoretical (Oshikawa {\it et al.} 1997, Totsuka 1998, Tonegawa 
{\it et al.} 1996, Cabra {\it et al.} 1997, Cabra and Grynberg 1999 
and Sakai and Takahashi 1998) 
and experimental (Narumi {\it et al.} 1998 and Shiramura {\it et al.} 1998) 
studies suggested the 
realization of the magnetization plateaux in various systems. 

The previous works (Yamamoto and Sakai 1998 and Sakai and Yamamoto 1998) 
by the present authors indicated 
an important role of the quantum fluctuation to stabilize the plateau 
against the planar anisotropy in the ferrimagnetic chain. 
It results in the existence of the plateau even in the $XY$ model 
of the mixed spins 1, and 1/2. 
On the other hand, the classical mixed spin systems 
have the same plateau in the isotropic case. 
It implies that the plateau is originally based on 
a classical mechanism, although it is stabilized by a quantum effect. 
Thus it is difficult to say that the plateau symbolizes 
the quantum nature of the ferrimagnetic chains. 
In the present paper, other plateaux, essentially based on 
a {\it quantum} mechanism, are revealed to coexist with 
the above mentioned {\it classical} plateau, 
when both spins are lager than 1/2. 

\section{Quantum and classical plateaux}

The recent exact treatment (Oshikawa {\it et al.} 1997) for general quantum 
spin chains suggested that a magnetization plateau can 
appear based on the quantization of the magnetization under the condition, 
\begin{equation}
S_{\rm unit}-m={\rm integer},
\label{condition}
\end{equation}
where $S_{\rm unit}$ and $m$ are the total spin and 
magnetization per unit cell. 
Note that the relation (\ref{condition}) is only a necessary condition. 
Thus it doesn't guarantee the existence 
of the plateau nor specify any mechanisms 
of its formation. 
The condition (\ref{condition}) is still valid for the mixed spin chains 
of $S$ and $s$ ($S>s$), described by the Hamiltonian 
\begin{equation}
   {\cal H}
    ={\displaystyle\sum_{j=1}^N}
    \bigl[
     (1+\delta)
     (\mbox{\boldmath$S$}_{j}\cdot\mbox{\boldmath$s$}_{j})_\alpha
    +(1-\delta)
     (\mbox{\boldmath$s$}_{j}\cdot\mbox{\boldmath$S$}_{j+1})_\alpha
    -H(S_j^z+s_j^z)
    \bigr]\,,
   \label{ham}
\end{equation}
with $(\mbox{\boldmath$S$}\cdot\mbox{\boldmath$s$})_\alpha 
 =S^xs^x+S^ys^y+\alpha S^zs^z$. 
In the isotropic case ($\alpha=1$) with no bond alternation 
($\delta=0$), the system has a spontaneous magnetization $m_s\equiv S-s$.
Because of the antiferromagnetic gap the ground state with $m_s$ 
is so stable against the excitation increasing $m$ that there 
appears a magnetization plateau at $m=m_s$. (Kuramoto 1998) 
The previous works (Yamamoto and Sakai 1998 and Sakai and Yamamoto 1998) 
on the most quantized system 
($S,s$)=(1,1/2) suggested that the quantum fluctuation stabilizes 
the plateau against the $XY$-like anisotropy ($\alpha <1$) and 
the plateau phase extends to the Kosterlitz-Thouless phase boundary 
in the ferromagnetic region ($\alpha <0$). 
However, this plateau phase also includes the Ising limit 
($\alpha \rightarrow \infty$) without any other boundaries. 
Thus it is difficult to identify some quantum effects on the 
plateau formation, because it cannot be clearly distinguished from the 
Ising gap based on a classical mechanism. 
In fact the classical spin (vector) model with the same 
amplitudes ($S,s$)=(1,1/2) described by the same Heisenberg Hamiltonian 
(\ref{ham}) also has a plateau at $m=m_s$ for $\alpha =0$. 
Then the plateau at $m_s$ should be called a {\it classical plateau}. 
On the other hand, the condition of the quantization (\ref{condition}) 
suggests that 
some other plateaux can appear at higher magnetization 
$m=S-s+1, S-s+2, \cdots, S+s-1$ for $S>s>1/2$. 
These higher plateaux can never be explained 
by any classical mechanisms, because 
they cannot appear in the Ising model or classical Heisenberg model. 
Thus they should be called {\it quantum plateaux}, if they really appear. 
In the following sections, we perform some theoretical 
analyses for the systems of $(S,s)=(3/2,1)$ and (2,1) to 
justify the coexistence of the quantum 
and classical plateaux in the case of $S>s>1/2$.

\section{Low-lying excitations}

The optical mode of the low-lying excitations characterizes 
the feature of the initial plateau at $m_s$. 
In Fig. 1 we show the excitation spectra of 
the systems (a)(3/2,1) and (b)(2,1) for $\alpha =1$ and 
$\delta=0$, derived from the three methods; 
the quantum Monte Carlo simulation (QMC), 
the modified spin wave theory and the perturbation from the decoupled dimer. 
The first one gives the most precise results and the last one 
is based on the dimer state described as 
$\prod_j
 (A_j^\dagger)^{S-s}
 (A_j^\dagger b_j^\dagger-B_j^\dagger a_j^\dagger)^{2s}
 |0\rangle$,
making use of the Schwinger boson representation:
\begin{equation}
   \left.
   \begin{array}{ll}
      S_j^+=A_j^\dagger B_j\,,&
      S_j^z=\frac{1}{2}(A_j^\dagger A_j-B_j^\dagger B_j)\,,\\
      s_j^+=a_j^\dagger b_j\,,&
      s_j^z=\frac{1}{2}(a_j^\dagger a_j-b_j^\dagger b_j)\,,
   \end{array}
   \right.
   \label{boson}
\end{equation}
The excitation spectrum of each system has two branches characterizing 
the ferromagnetic (lower) and antiferromagnetic (upper) features, 
as well as the system (1,1/2). 
The calculated curves suggest that the spin wave  
is more suitable than the decoupled dimer to describe the 
behavior of the optical branch around its bottom ($k=0$). 
It implies that the classical picture (spin wave excitation 
from the N\'eel order) is more effective than the quantum one 
(dimer-breaking excitation) to explain 
the origin of the initial plateau, as expected 
in the above argument. 

\begin{figure}[htb]
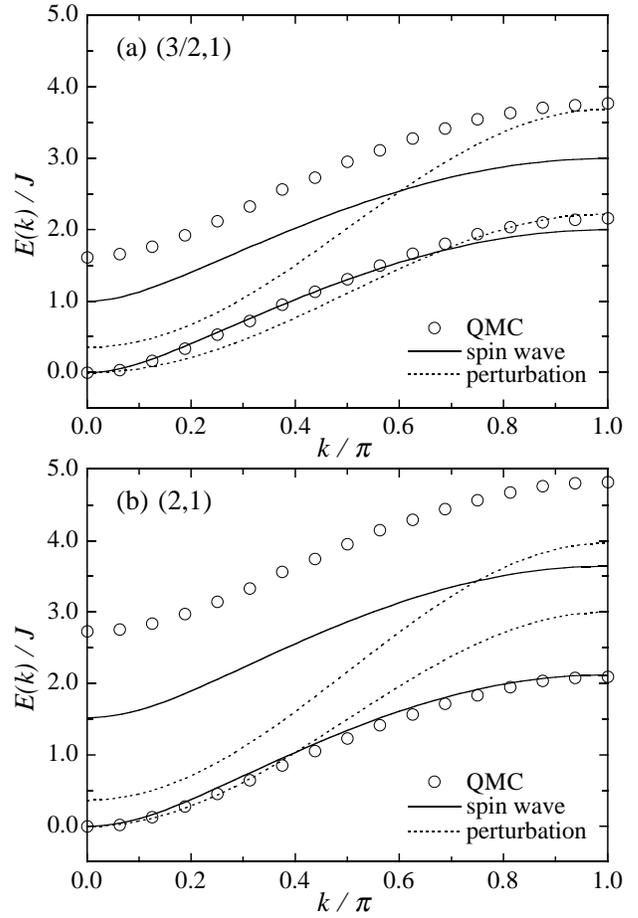

\begin{center}
\mbox{\psfig{figure=fig1a.epsi,width=8cm,height=6cm,angle=0}}
\mbox{\psfig{figure=fig1b.epsi,width=8cm,height=6cm,angle=0}}
\end{center}
\caption{
Low-lying excitation spectra calculated by the quantum Monte Carlo
simulation
(QMC), the spin wave theory and the perturbation from the decoupled
dimer for
(a)(3/2,1) and (b)(2,1).
\label{fig1}
}
\end{figure}

\section{Variational approach}

According to the condition of the quantization (\ref{condition}), 
the mixed-spin chains of (3/2,1) and (2,1) possibly have two 
plateaux at $m=m_s$ and $m_s+1$. 
In order to characterize these plateaux, we introduce a variational 
wave function for the ground state of the model (\ref{ham}) as 
\begin{eqnarray}
|{\rm g}\rangle 
=&c_{\rm N}\prod_{j=1}^N
(A_j^\dagger)^{2S}(b_j^\dagger)^{2s}|0\rangle \nonumber\\
&+ \sum_{l=0}^{2s}c_{\rm VB}^{(l)}
\prod_{j=1}^N
(A_j^\dagger)^{2S-l}(a_j^\dagger)^{2s-l}
(A_j^\dagger b_j^\dagger-B_j^\dagger a_j^\dagger)^l|0\rangle ,
\label{val}
\end{eqnarray}
where $c_{\rm N}$ and $c_{\rm VB}^{(l)}$ are the mixing coefficients. 
Using the variational wave function, the ground-state phase diagram in the 
$H$$\delta$ plane is obtained, as shown in Figs. 2 (a) and (b) 
for (3/2,1) and (2,1), respectively, where we restrict us on 
the Heisenberg point ($\alpha =1$). 
On the first step of the magnetization process for each system, there exists a 
crossover point $\delta _c$ between the N\'eel (N) 
and double-bond dimer (DBD) states. 
In contrast, the second step toward the saturation (S) 
is always the single-bond dimer 
(SBD) state. 
These two steps before the saturation are expected to characterize 
the two plateaux. 
Thus the first plateau should be based on the classical N\'eel order, 
while the second one on the quantum Valence-Bond-Solid state, 
as far as we consider the case of small $\delta$. 

\begin{figure}[htb]
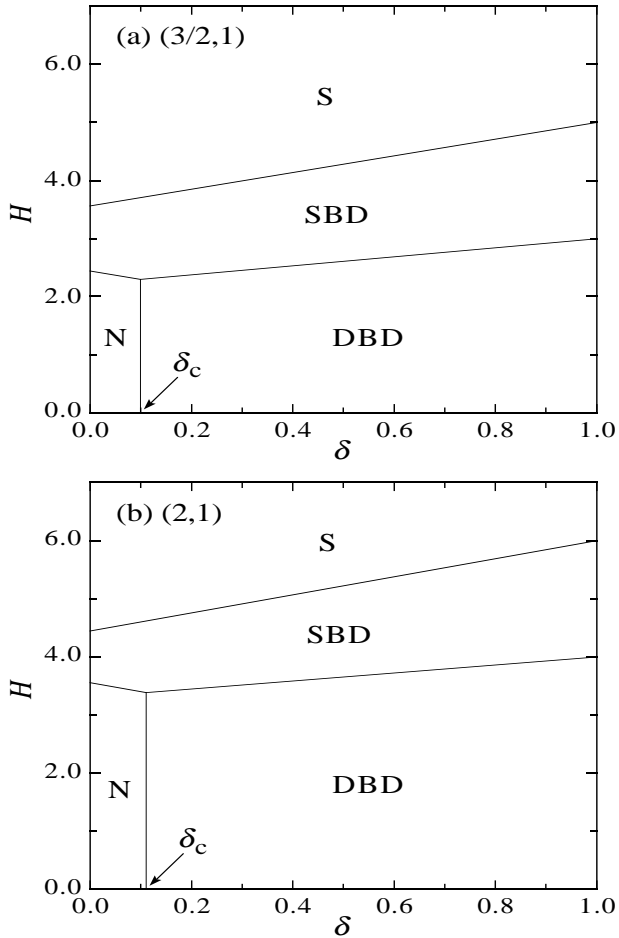

\begin{center}
\mbox{\psfig{figure=fig2a.epsi,width=8cm,height=6cm,angle=0}}
\vskip 3mm
\mbox{\psfig{figure=fig2b.epsi,width=8cm,height=6cm,angle=0}}
\end{center}
\caption{
The variational ground-state phase diagrams on the
$\delta H$-plane with $\alpha=1$ for (a)(3/2,1) and (b)(2,1).
Phases are denoted as N: N\'eel, DBD: double-bond dimer,
SBD: single-bond dimer and S: saturation, respectively.
\label{fig2}
}
\end{figure}

\section{Phase diagrams}

In order to confirm the coexistence of the two plateaux 
even for $\alpha =1$ and $\delta =0$, we perform a size scaling 
analysis with the exact diagonalization of finite systems 
up to $N=12$ to present the phase diagrams in the $\delta$$\alpha$ plane. 
$E(N,M)$ denotes the lowest energy in the subspace with a fixed 
magnetization $M$ for the Hamiltonian (\ref{ham}) without the Zeeman term. 
The upper and lower bounds of the external field which induces the 
ground-state magnetization $M$ are expressed as 
$H_\pm(N,M)=\pm E(N,M\pm 1)\mp E(N,M)$. 
The length of the plateau with the unit-cell magnetization 
$m\equiv M/N$ is obtained as ${\mit\Delta}_N(m)=H_+(N,M)-H_-(N,M)$. 
The quantity $\Delta _N(m)$ also corresponds to the sum of the 
two excitation gaps for increasing and reducing the magnetization, 
respectively.(Sakai and Takahashi 1998)  
Thus the scaled quantity $N{\mit\Delta}_N(m)$ is a good probe of the plateau. 
No size dependence means that the system is gapless. 
The scaled quantity 
of the system (3/2,1) at (a)$m=1/2$ and (b)$m=3/2$ is shown as 
a function of $\alpha$ for $\delta=0$ in Fig. 3. 
Fig. 3 (a) clearly shows that opening plateau around 
$\alpha=1$ vanishes at some critical value $\alpha _c$ for $m=1/2$ 
and a gapless phase lies in the region of $\alpha < \alpha _c$. 
The scaled gap in Fig. 3 (b) also indicates the existence 
of the second plateau around $\alpha=1$, although the size 
dependence is much smaller than that of the first plateau. 
It implies that the second plateau is much smaller than the first one. 

\begin{figure}[htb]
\begin{center}
\mbox{\psfig{figure=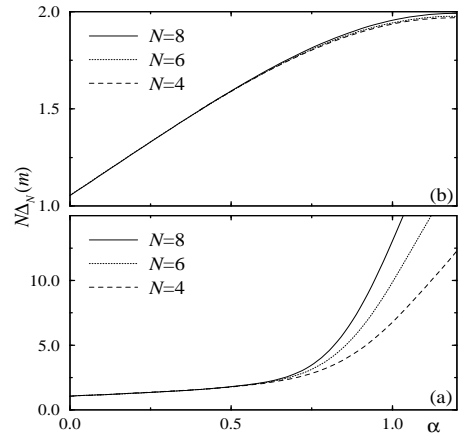,width=8cm,height=6cm,angle=-90}}
\end{center}
\caption{
Scaled quantity $N{\mit\Delta}_N(m)$ versus $\alpha$ at
$m=\frac{1}{2}$ (a) and $m=\frac{3}{2}$ (b) in the case of
$(S,s)=(\frac{3}{2},1)$.
\label{fig3}
}
\end{figure}

To investigate the critical property around $\alpha _c$, 
we use the size scaling formula based on the conformal field theory 
(Cardy 1984, Bl\"ote {\it et al.} 1986 and Affleck 1986) 
\begin{equation}
   \frac{1}{N}E(N,M)
    \sim \epsilon (m) - \frac{\pi v_{\rm s}c}{6N^2}\,,
    \label{gs}
\end{equation}
and
\begin{equation}
   {\mit\Delta}_N(m)
    \sim \frac{\pi v_{\rm s}\eta}{N}\,,
    \label{gap}
\end{equation}
with the following notations: 
$\epsilon (m)$ is the ground state energy of the bulk system; 
$v_{\rm s}$ is the sound velocity derived from the derivative 
of the dispersion curve at $k=0$; 
$c$ is the central charge; 
$\eta$ is the critical exponent defined by the spin-correlation function 
$\langle \tilde s^x_0 \tilde s^x_r \rangle \sim r^{-\eta}$, 
where $\tilde {\bf s}_i$ is some 
relevant spin operator. 
They are valid at gapless points. 
The calculated $c$ and $\eta$ of the systems (3/2,1) and (2,1) at $m=m_s$ and 
$m=m_s+1$ indicate the following properties: 
as $\alpha $ decreases from 1 with fixed 
$\delta$, the first and second plateaux vanish 
at the different Kosterlitz-Thouless 
critical points $\alpha _{c1}$ and $\alpha _{c2}$, 
respectively, where $\eta $ is 
1/4 in common; 
the gapless spin fluid phase characterized by $c=1$ lies in the region 
$\alpha <\alpha _{c1}$ ($\alpha _{c2}$) at $m=m_s$ ($m=m_s+1$). 
The universality of the phase boundary 
of both plateaux are the same as that of the 
unique plateau of the system (1,1/2) 
(Yamamoto and Sakai 1999 and Sakai and Yamamoto 
1999). 
Thus we determine the gapless-plateau phase boundary 
by $\eta =1/4$ for both plateaux. 
Thus-obtained phase boundaries for the first and second plateaux 
are shown together as solid lines in Figs. 4 (a) and (b) 
for the systems (3/2,1) and (2,1), respectively. 
The phase diagrams obviously turn out the coexistence of the 
first (classical) and second (quantum) plateaux even at the 
most symmetric point ($\alpha =1$ and $\delta =0$). 
They also exhibit an interesting feature; the quantum plateau 
phase is larger than the classical one 
($\alpha _{c1}\leq \alpha _{c2}$ independently 
of $\delta$). 
It implies that the quantum plateau is more stable than 
the classical one against the planar anisotropy. 

\begin{figure}[htb]
\begin{center}
\mbox{\psfig{figure=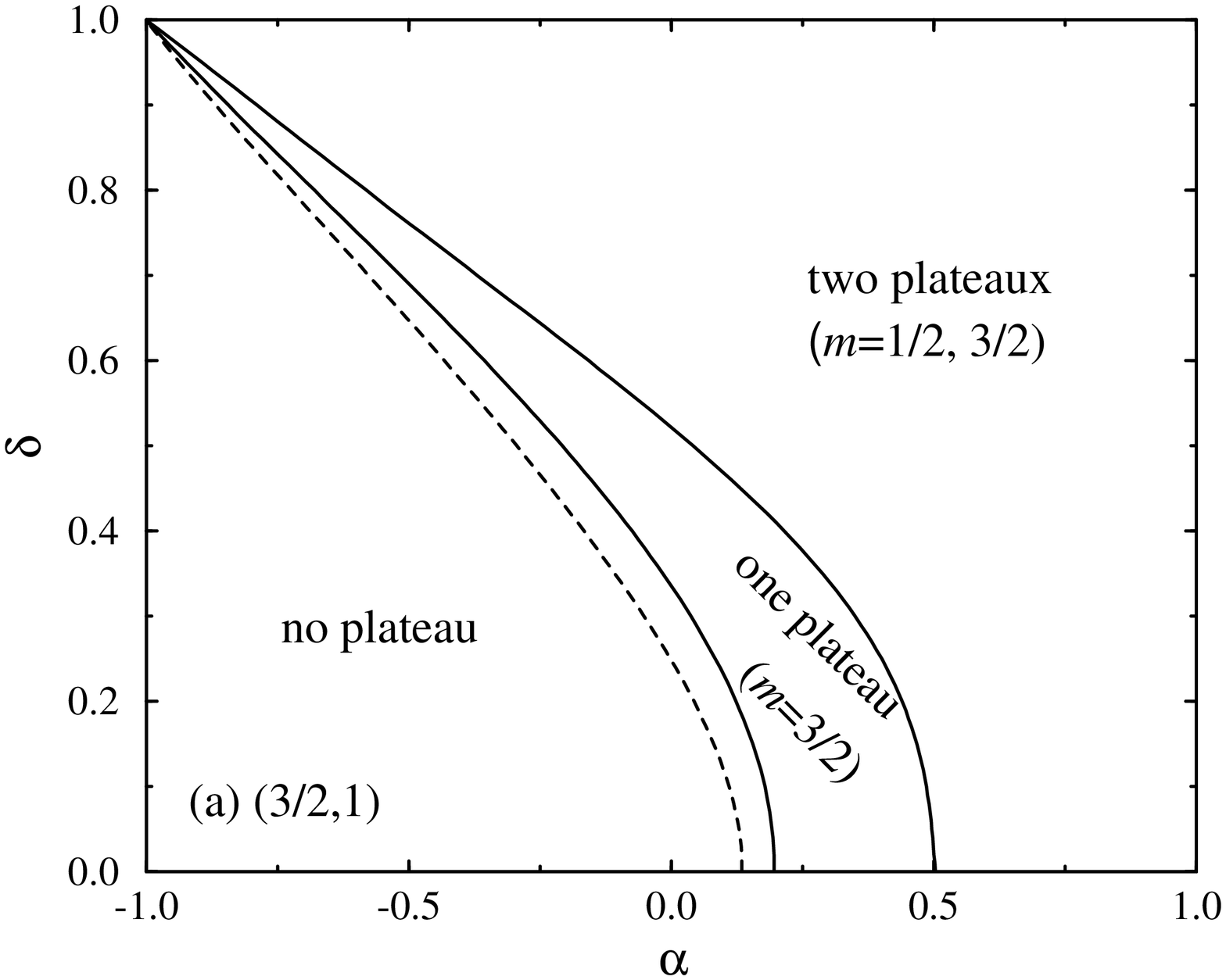,width=8cm,height=6cm,angle=-90}}
\mbox{\psfig{figure=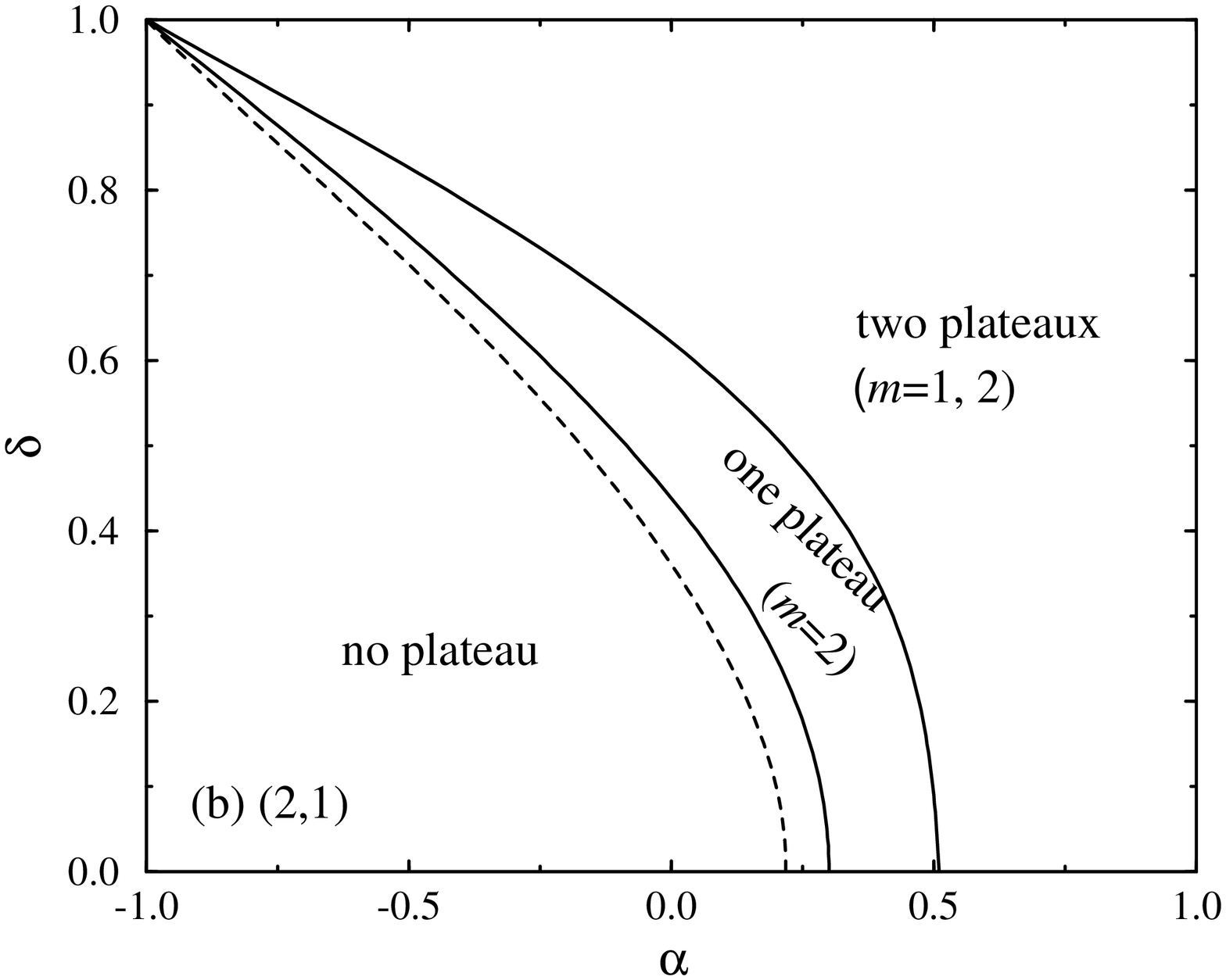,width=8cm,height=6cm,angle=-90}}
\end{center}
\caption{
The ground-state phase diagrams on the $\alpha\delta$-plane
for (a)(3/2,1) and (b)(2,1).
Solid lines are the phase boundaries determined by $\eta =1/4$.
Dashed lines are the boundaries of the second plateaux determined
by the level spectroscopy.
A small difference between the two results for the second plateaux
is due to the logarithmic size correction which should appear
in the former method.
\label{fig4}
}
\end{figure}

Such a slight size dependence of the scaled quantity 
$N\Delta _N(m)$ for the second 
plateau as shown in Fig. 3 (b) might make us doubt 
its existence for $\alpha =1$ and $\delta =0$. 
Then we perform another analysis, 
called level spectroscopy (Okamoto and Nomura 1992 and Nomura 1995), 
to convince of the existence of the second plateau. 
It is one of the most precise methods to estimate 
the Kosterlitz-Thouless phase boundary. 
Since the method detects the boundary 
as a level crossing point of the two relevant 
excitation gaps with the same scaling dimension, 
the result does not suffer from the 
dominant logarithmic size correction, which is quite serious for 
the Kosterlitz-Thouless transition. 
For the second plateau at $m=S-s+1$, the two relevant gaps are given by 
\begin{equation}
\Delta _0 \equiv E_2(L,M_2)-E(L,M_2) , 
\label{gap0}
\end{equation}
\begin{equation}
\Delta _4
\equiv [E(L,M_2+4)+E(L,M_2-4)-2E(L,M_2)]/2,
\label{gap4}
\end{equation}
where $E_2(L,M)$ is the second eigenvalue in the same subspace 
as E(L,M) and $M_2$ is defined as $M_2\equiv (S-s+1)N$. 
The two excitations have a common scaling dimension 2. 
The calculated $\Delta _0$ and $\Delta _4$ of 
the system (3/2,1) for $\delta =0$ are 
plotted versus $\alpha $ in Fig. 5. 
It suggests that the phase boundary is easily 
determined as a crossing point, almost 
independent of the system size. 
Thus-obtained boundaries for the second plateaux 
are shown as dashed curves in Figs. 
4 (a) and (b). 
The results have a little deviation from 
the boundaries by $\eta =1/4$, because the 
latter includes the logarithmic size correction. 
Anyway they lead to the same conclusion; 
the coexistence of the classical and quantum 
plateaux for $\alpha =1$ and $\delta =0$. 

\begin{figure}[htb]
\begin{center}
\mbox{\psfig{figure=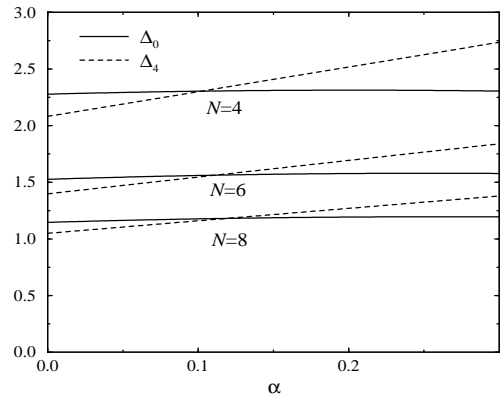,width=8cm,height=6cm,angle=-90}}
\end{center}
\caption{
$\Delta _0$ and $\Delta _4$ of the system (3/2,1) with
$\delta=0$ versus $\alpha$.
It indicates that the crossing point of the two gaps
is almost independent of the
system size.
\label{fig5}
}
\end{figure}

The N\'eel-dimer crossover point 
in the first plateau indicated by the variational 
method in the previous section is not detected as any phase boundaries by these 
numerical analyses. 
It suggests that the N\'eel and 
dimer pictures cannot be distinguished clearly for 
the first plateau. 
In fact it is trivially revealed that in the $\delta$$\alpha$ phase diagram at 
$m=m_s$ the isotropic dimer point 
($\alpha=1$ and $\delta=1$) is connected to the 
Ising limit ($\alpha \rightarrow \infty$ and $\delta=0$) 
via the Ising dimer limit 
($\alpha \rightarrow \infty$ and $\delta=1$) through no phase transition or 
crossover. 
It implies that the first plateau always 
bears an aspect of the Ising gap even for large $\delta$. 

\begin{figure}[htb]
\begin{center}
\mbox{\psfig{figure=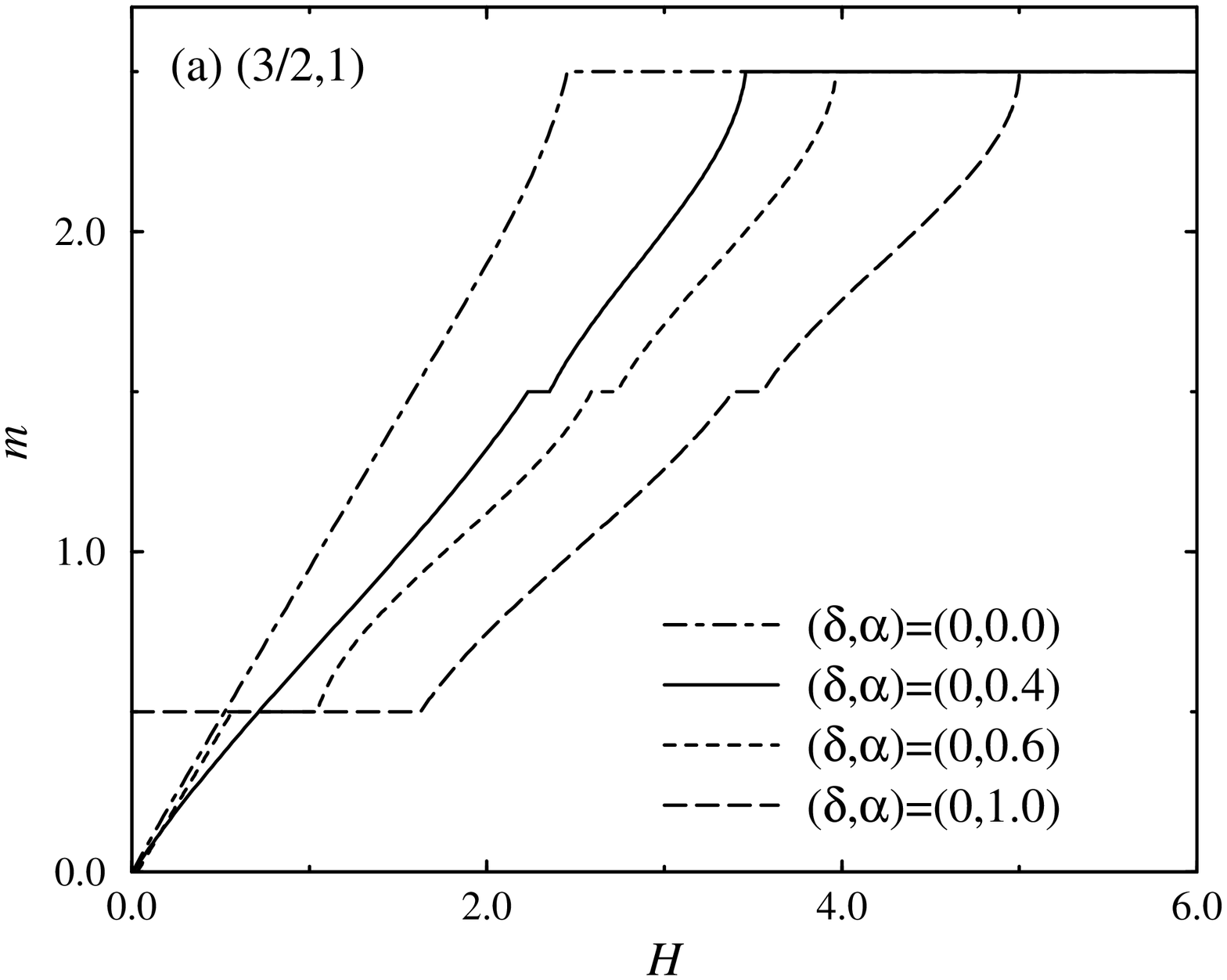,width=8cm,height=6cm,angle=-90}}
\mbox{\psfig{figure=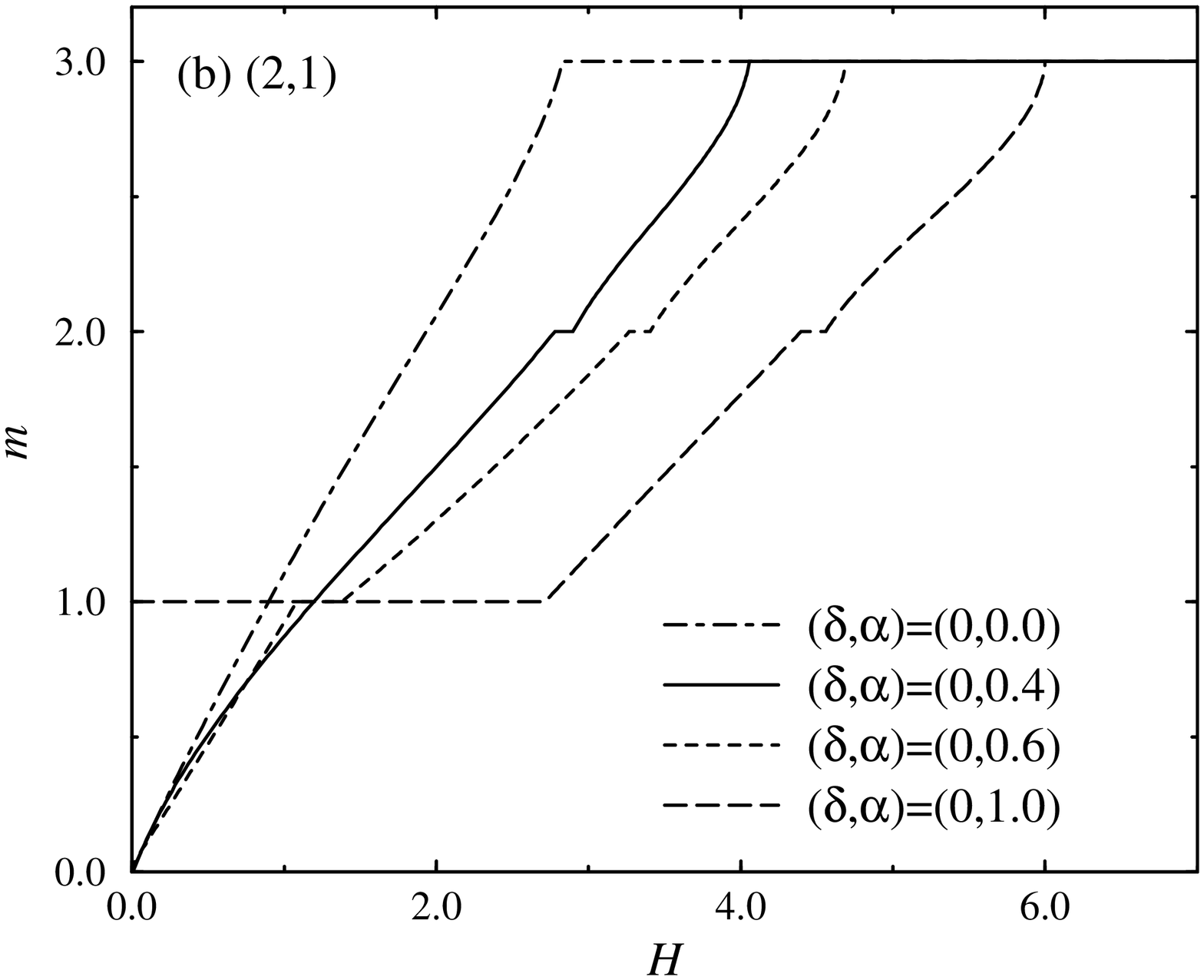,width=8cm,height=6cm,angle=-90}}
\end{center}
\caption{
The ground-state magnetization curves of the quantum system
with $\delta=0$ at various values of $\alpha$
for (a)(3/2,1) and (b)(2,1).
\label{fig6}
}
\end{figure}

\section{Magnetization curve}

Finally we present the ground state magnetization curve in several cases of the 
systems (3/2,1) and (2,1). 
The curve is given by extrapolating 
$H_\pm(N,M)$ to the thermodynamic limit using 
the size scaling 
(Sakai and Takahashi 1998) based on the conformal field theory at 
gapless points and the Shanks transformation for plateaux. 
We show only the results of 
a suitable polynomial fitting to thus-obtained points, 
in Figs. 6 ($\delta =0$) 
and 7 ($\delta=0.4$), where the labels 
(a) and (b) indicate the systems (3/2,1) and (2,1), respectively. 
They visualize the coexistence 
of the classical and quantum plateaux at the Heisenberg 
point. 
These results also justify 
the above mentioned feature; the quantum plateau is smaller 
at the Heisenberg point 
but more stable against the $XY$-like anisotropy, rather than 
the cassical one. 
Therefore, 
if these systems lose a plateau due to the anisotropy, only the quantum 
one would survive. 
In order to clarify the difference 
in the mechanism of the gap formation between the 
first and second plateaux, 
we also present the magnetization curves of the classical 
Heisenberg spin systems described 
by the same Hamiltonian (\ref{ham}) with the same 
amplitudes $(S,s)=(3/2,1)$ 
and (2,1) in Figs. 8 (a) and (b), respectively. 
(The results are independent of $\delta$.) 
The classical systems clearly have 
the first plateau in the isotropic case, while 
there appears no plateau corresponding to the second one. 
It also supports the quantum nature of the second plateau. 
These plateaux of the classical systems vanishes even for a slight anisotropy; 
$\alpha _c=0.980$ and 0.943 for $(S,s)$=(3/2,1) and (2,1), repectively. 
In comparison with these critical values, 
the phase boundaries in the quantum systems in Figs. 4 (a) and (b) 
imply that quantum fluctuation stabilies even the first plateau. 
Thus, in geneal, 
the quantum effect is expected to toughen every field-induced gap 
against the planar anisotropy. 
Nevertheless the first and second plateaux 
should be distinguished, because the 
former appears even in the classical limit, 
while the latter doesn't exist until the 
spin is quantized. 

\begin{figure}[htb]
\begin{center}
\mbox{\psfig{figure=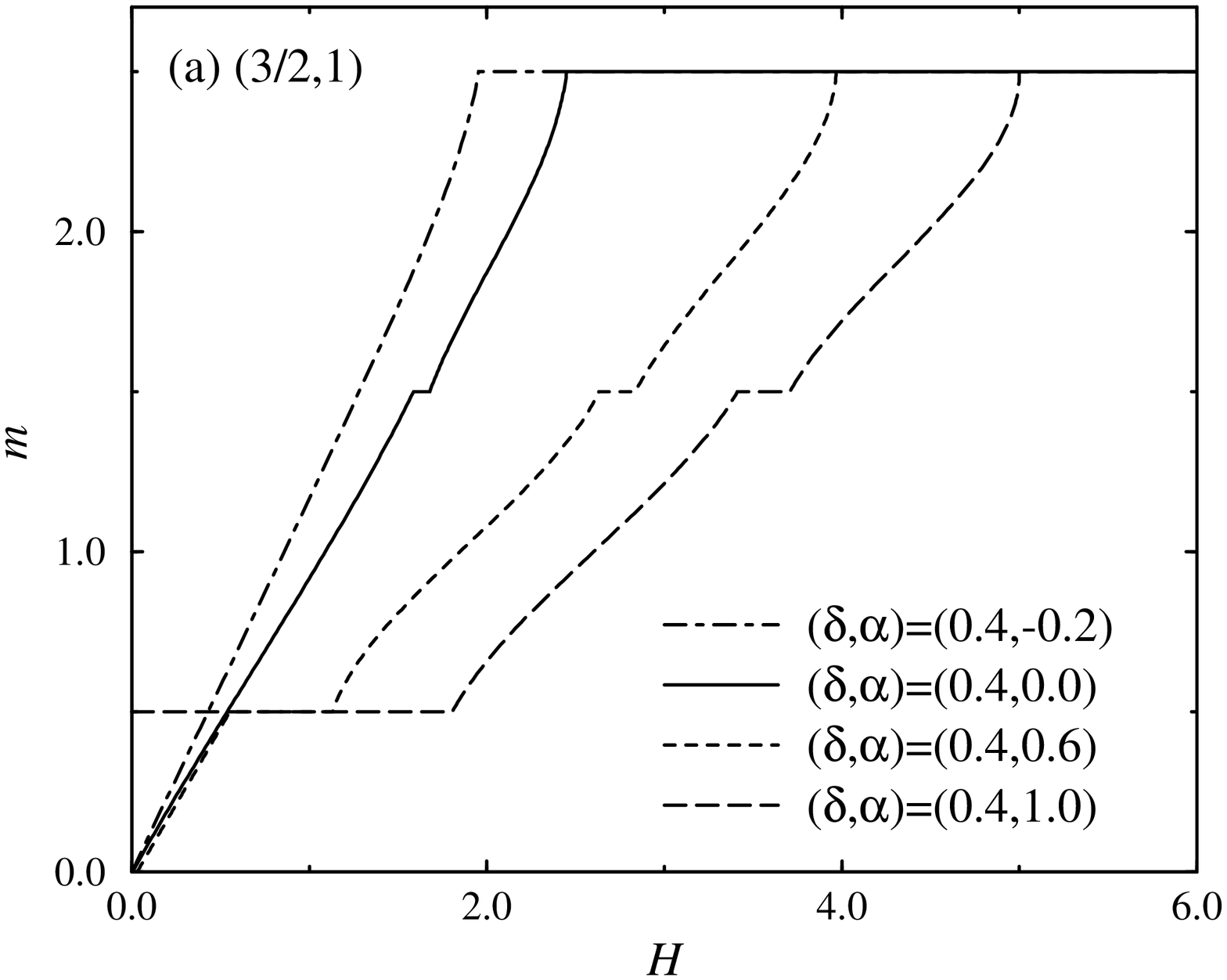,width=8cm,height=6cm,angle=-90}}
\mbox{\psfig{figure=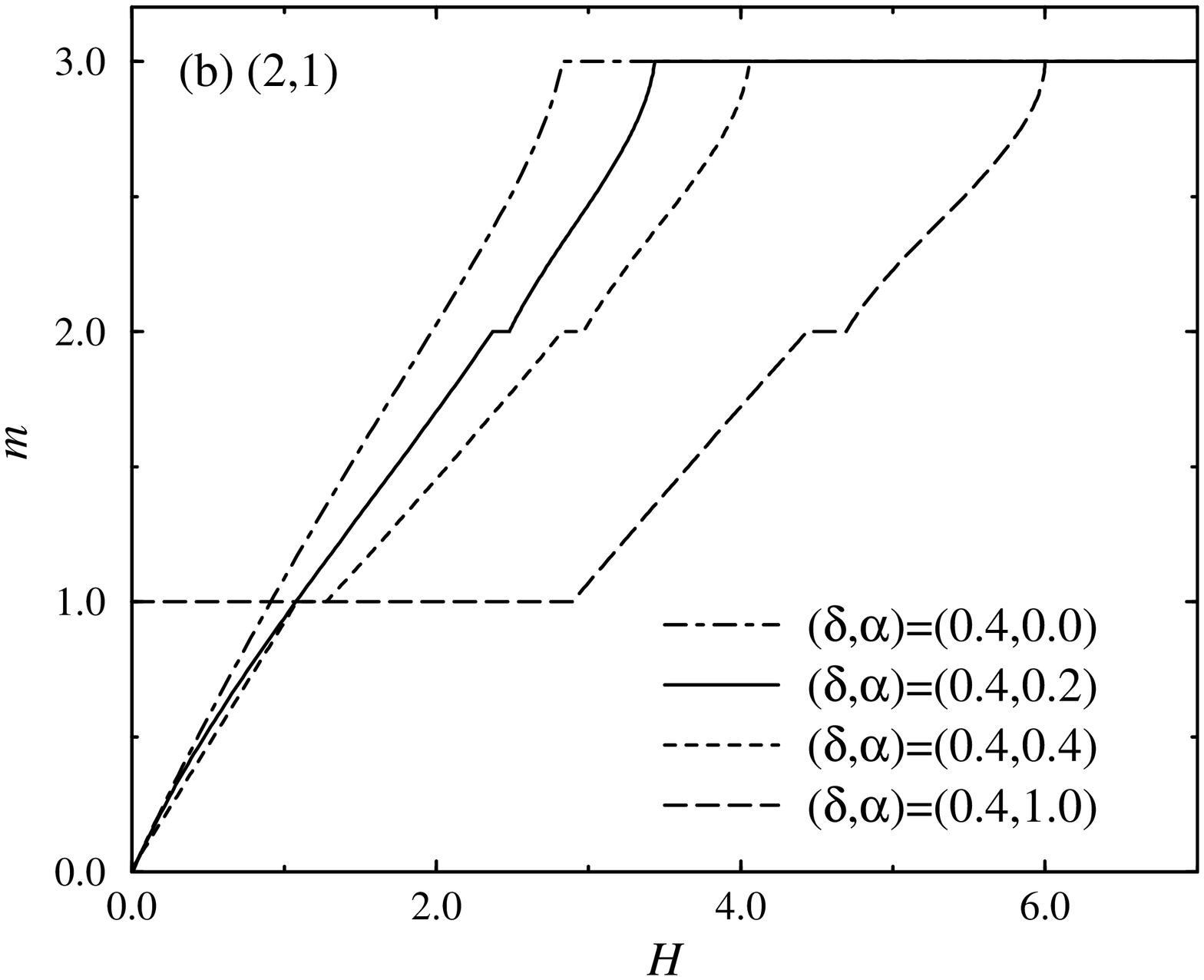,width=8cm,height=6cm,angle=-90}}
\end{center}
\caption{
The ground-state magnetization curves of the quantum system
with $\delta=0.4$ at various values of
$\alpha$ for (a)(3/2,1) and (b)(2,1).
\label{fig7}
}
\end{figure}

\begin{figure}[htb]
\begin{center}
\mbox{\psfig{figure=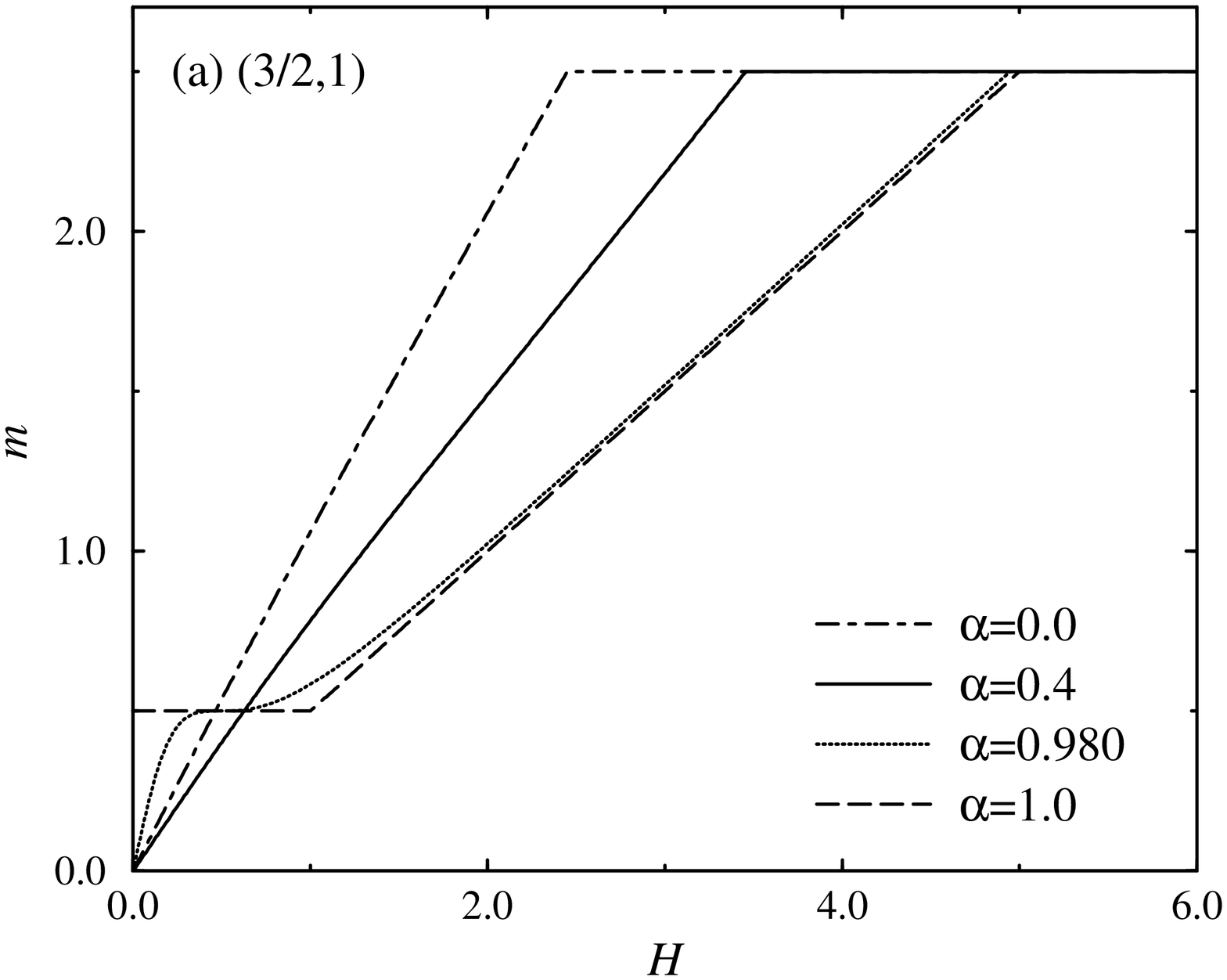,width=8cm,height=6cm,angle=-90}}
\mbox{\psfig{figure=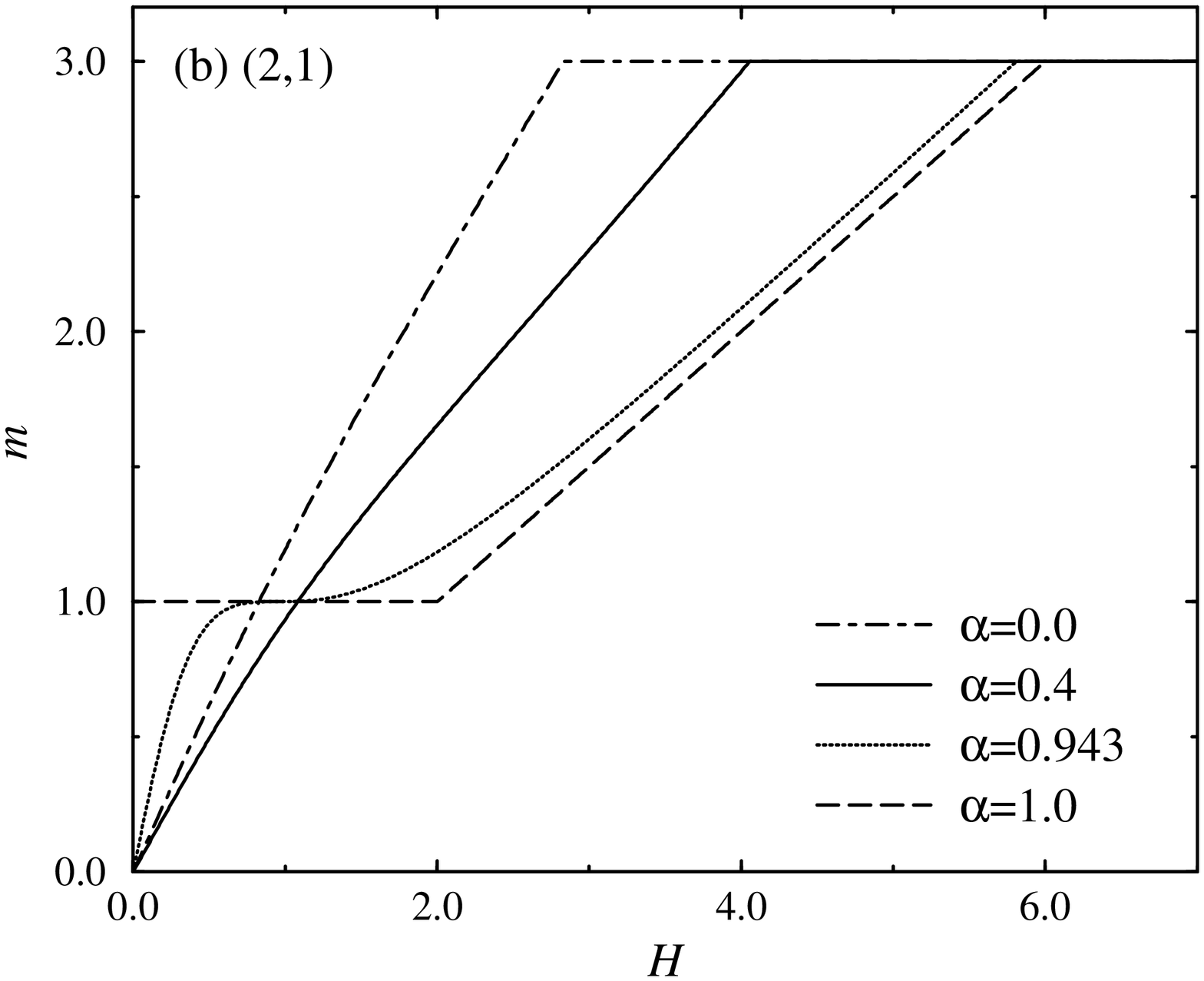,width=8cm,height=6cm,angle=-90}}
\end{center}
\caption{
The ground-state magnetization curves of the classical system
at various values of $\alpha$ for (a)(3/2,1) and (b)(2,1).
\label{fig8}
}
\end{figure}

\section{Concluding remarks}

The above investigations turn out the coexistence of the classical 
and quantum plateaux in the ground-state magnetization curve of 
the mixed spin chains of (3/2,1) and (2,1). 
The conclusion is easily generalized for (S,s) ($S>s>1/2$); 
the magnetization curve has $2s$ plateaux and only the initial one 
is based on a classical mechanism, while the other ones originate in 
quantum correlations. 

In most previous works 
on the magnetizaion plateau the gap formation were based on 
the bond polymerization. 
(Totsuka 1998, Tonegawa {\it et al.} 1996, Cabra {\it et al.} 1997, Cabra 
and Grynberg 1999) 
In contast, the present proposal of the plateau in ferrimagnetic chains is a 
pioneering trial to explore 
a novel mechanism of the field-induced gap associated 
to the {\it spin polymerization}. 
The bimetallic chains such as 
MM$'$(pbaOH)(H$_2$O)$_3\cdot$nH$_2$O (Karn 1989 and 
Kahn {\it et al.} 1995) are good candidates to realize the spin plymerization. 
Unfortunately, most of them are in the case of M$'$=Cu, that is $s=1/2$. 
In fact a few compounds with other metals were also synthesized, for example, 
MM$'$(EDTA)6H$_2$O (MM$'$=CoNi, MnCo and MnNi). 
However, 
the case of MM$'$=CoNi yields $(S,s)$=(1(Ni),1/2(Co)) (Drillon {\it et
al.} 1985) 
and MnCo a large Ising-like anisotropy. (Drillon {\it et al.} 1986) 
Thus they are not any suitable stages to search for the quantum plateau. 
Among the series of bimetallic chains, the most suitable compound might be 
MnNi(EDTA)6H$_2$O, 
which is well described by the 1D spin-alternating Heisenberg 
model of (5/2,1). (Drillon {\it et al.} 1986) 
The magnetization measurement on it 
would be interesting to investigate a possible 
quantum plateau at $m=5/2$, as well as the classical one at $m=3/2$. 

One of the most important remarks 
in the present work is the coexistence of the quantum 
and classical plateaux even 
at the most symmetric point ($\alpha=1$ and $\delta =0$). 
In order to examine it, 
compounds consisting of metals and stable organic radicals 
(Caneschi {\it et al.} 1989 and Markosyan {\it et al.} 1998) 
might be a more ideal stage, because 
the organic radicals lead to entirely isotropic spin systems, rather than the 
bimetallic chains with inevitable Ising-like anisotropy. 
The metal-radical complex also has a lot of variations. 
The recently synthesized one 
\{Mn(hfac)$_2$\}$_3$(3R)$_2$ (Markosyan {\it et al.} 1998) has 
been investigated to realize the (5/2,3/2) spin chain. 
We hope the present calculations will stimulate not only further theoretical 
investigations, but also experimental 
explorations into the magnetization plateaux 
in ferrimagnets as spin-polymerized materials. 

\section*{Acknowledgment}  
The authors thank Dr. K. Okamoto for useful discussion. 
This work is supported by the Japanese Ministry of Education, 
Science, and Culture through Grant-in-Aid No. 11740206 and by the 
Sanyo-Broadcasting Foundation for Science and Culture. 
The computation was done in part using the facility of the 
Supercomputer Center, Institute for Solid State Physics, University 
of Tokyo.


%


%

\widetext
\end{document}